\documentclass{kapproc} 

\RequirePackage{graphicx}%
\RequirePackage{epsf}%
\input{psfig}

\upperandlowercase
\let\footnote\savefootnote
\let\footnotetext\savefootnotetext 
 
\kluwerbib 

\def\msun{M$_\odot$\ }
\def\msunn{M$_\odot$}

\begin{document}

\articletitle{The 0.03--10\msun mass function of young open clusters}
\chaptitlerunninghead{The mass function of young open clusters}

 \author{J. Bouvier\altaffilmark{1}, E. Moraux\altaffilmark{2}, and J. Stauffer\altaffilmark{3}}
 \affil{\altaffilmark{1}LAOG, Grenoble, France\\ 
\altaffilmark{2}IoA Cambridge, UK\\
\altaffilmark{3}IPAC, Caltech, USA }


 \begin{abstract}
   We report the present day mass functions (PDMFs) of 3 young open
   clusters over a mass range from 30 Jupiter masses to 10~\msunn. The PDMFs
   of the 3 clusters are remarkably similar, suggesting little impact of
   specific conditions (stellar density, metallicity, early dynamical
   evolution) on the mass distribution. Functional forms are provided to
   allow quantitative comparison with MFs derived in other environments.

 \end{abstract}

\section{Introduction}

Now that hundreds of brown dwarfs have been discovered, one can start to
address quantitatively such issues as the continuity of the mass function
across the stellar/substellar boundary, the overall shape of the IMF over
several decades of mass, its invariance or, on the contrary, its dependency
on local conditions (metallicity, stellar density, ionizing flux, etc.),
and to search for the lower mass cutoff corresponding to the fragmentation
limit below which a condensed object cannot form in isolation. Such a
complete characterization of the IMF would provide definite constraints on
theories of star and brown dwarf formation. 


The determination of bias-corrected mass functions over nearly 3 decades of
masses, from 0.03 to 10\msunn, is presented in this contribution for young
open clusters (YOCs). The benefits of young clusters and the resulting
reliability of the derived MF are outlined in Section~2. Deep wide-field
optical surveys performed at CFHT uncovered several tens of brown dwarfs in
3 nearby YOCs and are briefly presented in Section~3. The resulting mass
functions we derive for the 3 clusters are presented in Section 4 and
analytical fits are obtained, which permit quantitative comparison with
model predictions and/or mass distributions derived in other environments
(SFRs, field, extragalactic).

\section{The promise of young open clusters }

Deriving a reliable estimate of the mass function especially in the
substellar domain is an intrincate multistep process. A proper sample of
confirmed brown dwarfs must first be built, usually from an initial
photometric selection of candidates followed by a spectroscopic assessment
of their substellar nature. The distribution of absolute magnitudes (or
luminosities) is derived, provided that the distance to each object is
known and a correction for foreground extinction is made. Then, the
knowledge of the age of each object is required to convert its luminosity
to a mass, with the help of age-dependent mass-luminosity relationships
delivered by evolutionary models. The reliability of the resulting mass
distribution thus largely depends on the precision attached to the
distance, age and extinction of each object in the sample. Uncertainties in
any of these quantities will most likely skew the resulting mass function,
as will any sytematic error in the theoretical models. In addition, scatter
in the mass function estimate may result from low number statistics, which
can only be beaten by the construction of large samples, and by intrinsic
variations in the properties of the objects in the sample, such as age or
metallicity.

In young open clusters, all these sources of uncertainties are reduced.
Rich nearby clusters harbour large populations of stars and brown dwarfs
whose distance and age are known, usually to better than 20\% and sometimes
to within a few percent. Extinction is most often insignificant and scatter
in the intrinsic properties of the cluster members (age, distance,
metallicity) minimal. Spectroscopic diagnostics of youth, such as lithium
or gravity sensitive features, serve to distinguish between bona fide
substellar cluster members and intervening field dwarfs. A property
specific to a nearby young cluster, the common proper motion shared by all
its members also offers an additional criterion to build homogeneous
samples of confirmed cluster members. Hence, with distance and age
accurately known, and membership assessed in several independent ways,
observed magnitudes of cluster members are directly converted to mass with
the help of theoretical models. The resulting mass function pertains to an
homogenous population of objects spanning several decades of mass from low
mass brown dwarfs to the most massive cluster members, all of which were
born together under the same conditions and have shared the same evolution.

Another advantage is the concentration of cluster populations over
restricted areas on the sky, which can now be mapped in a very efficient
way with the new generation of wide-field CCD mosaics. Owing to their youth
and proximity, substellar cluster members are still relatively bright and
easily detectable from intermediate size telescopes.  Hence, deep large
scale surveys can be devised which cover a large fraction of the cluster
area, as is required to characterize the spatial distribution of cluster
members as a function of mass. How the cluster's present day mass function
(PDMF) relates to the IMF is, in principle, a matter of concern. However,
numerical simulations suggest that it is not until a few 100 Myr that
clusters start to preferentially lose low mass members as they relax
dynamically. By 100 Myr, the typical age of the clusters studied in the
next section, only 10\% of the stars (and brown dwarfs) are expected to
have been removed from the cluster (e.g. de la Fuente Marcos \& de la
Fuente Marcos 2000).  Primordial ejection of brown dwarfs from their
protostellar cradle, as advocated by Reipurth \& Clarke (2001), does not
appear to be an issue either since recent numerical simulations indicate
that ejection velocities are similar for brown dwarfs and stars, and are
usually smaller than the velocity required to escape the cluster's
potential (e.g.  Delgado, Clarke \& Bate 2004, Moraux \& Clarke 2004).

Thus, young open clusters offer a ground to build statistically robust mass
functions from volume limited and physically homogeneous samples of coeval
stars and brown dwarfs. At an age of about 100 Myr, the derived mass
function is predicted to be quite similar to the IMF, i.e., prior to any
dynamical evolution of the cluster.  Due to the distance of the clusters,
multiple systems are unresolved. {\it System} mass functions are thus
derived, which can nevertheless be statistically corrected for binarity to
yield single star (and brown dwarf) mass distributions (see, e.g., Moraux
et al. 2003).

\section{Brown dwarfs in young open clusters}

We performed deep, wide area photometric surveys at CFHT of 3 nearby open
clusters whose properties are summarized in Table 1. Beyond their proximity
and richness, they have also been selected as having a similar age, between
100 and 150 Myr, but possibly differing in metallicity. The photometric
survey covers a large enough area in each cluster so as to derive the
spatial distribution of cluster members as a function of mass and hence
take spatial segregation into account when estimating the cluster's mass
function. The depth of the survey (I$\sim$z$\sim$24.0) corresponds to a
mass limit between 30 and 50 Jupiter masses. On the bright side, saturation
occurs on the images for $\sim$0.4 \msun stars. Mass functions are therefore
derived in the mass range 0.030-0.40\msunn, and completed by literature
data for more massive cluster members. Binaries and higher order multiple
systems are not resolved at the distance of the clusters.

\begin{table}[ht]
\caption{Deep wide-field surveys of young open clusters.}
\begin{tabular*}{\textwidth}{@{\extracolsep{\fill}}lccccc}
\sphline
\it Cluster&\it Age &\it [Fe/H] &\it Richness &\it Distance & \it Surveyed area \cr
& \it (Myr) && \it (known members) & \it (pc) & \it (sq.deg.) \cr
\sphline
Blanco 1 & 100 & +0.1/0.2 & $>$ 200 & 260 & 2.5 \cr
Pleiades & 120 & 0.0      & $\sim$ 1200 & 130 & 6.4 \cr 
NGC 2516 & 150 & -0.3/0.0 & $\sim$ 2000 & 350 & 2.0 \cr
\sphline
\end{tabular*}
%
\end{table}

\begin{figure}[ht]
\sidebyside
{\centerline{\psfig{file=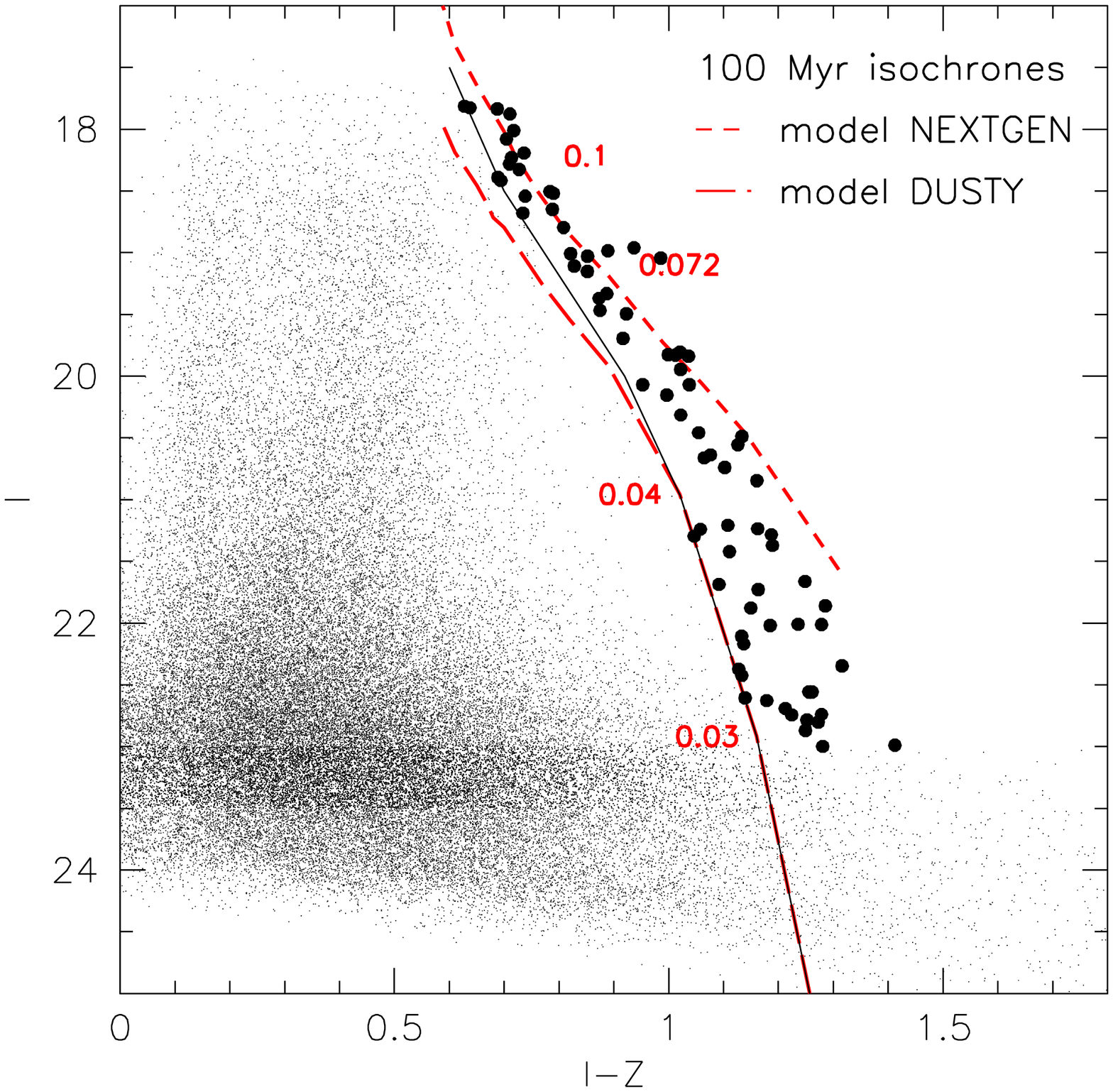,height=2.7in}}
\caption{Optical CMD of the Blanco 1 cluster : 100 Myr
  NextGen and Dusty isochrones (dashed) from the Lyon group are labelled
  with mass.  The cluster sequence, on the red side of our selection line
  (solid), is shown by filled dots. (From Moraux et al.  2004). }}
{\centerline{\psfig{file=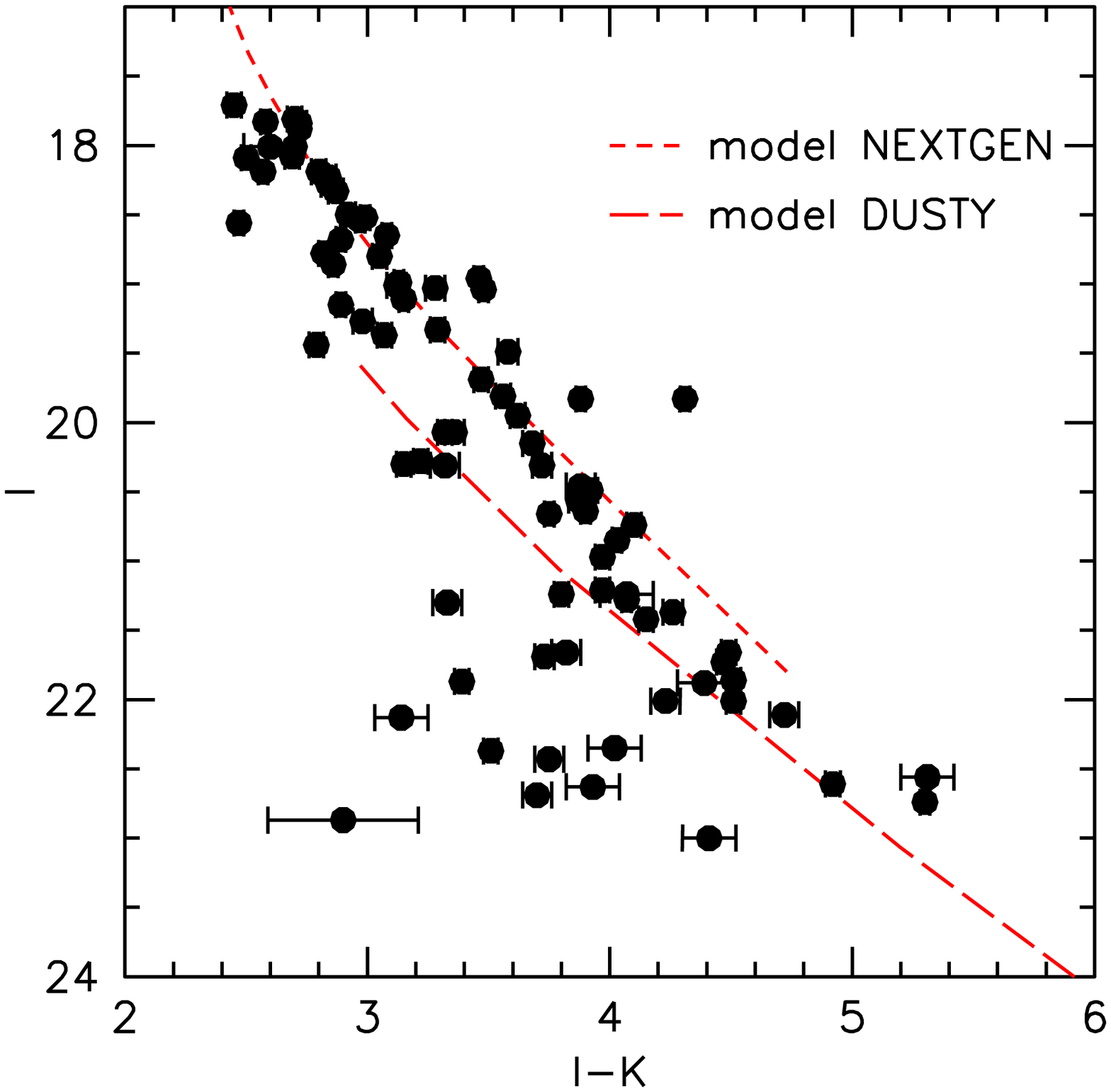,height=2.6in}}
\label{blanco1a}
\caption{Infrared  CMD of the Blanco 1 cluster
  candidates : optically-selected candidates lying on the blue side of the
  isochrones in this diagram are rejected as non members. (From Moraux et
  al. 2004). }}
\label{blanco1b}
\end{figure}

The photometric selection in (I, I-z) color magnitude diagrams yields from
several tens to several hundred low mass member candidates in each cluster
(e.g. Moraux et al.  2003). The candidates were followed up and their
membership assessed (or rejected) based on a combination of diagnotics~:
proper motion, lithium absorption, spectroscopic gravity index, infrared
photometry and/or spectroscopy. As an illustration of the method and the
results, Figures~\ref{blanco1a}~\&~\ref{blanco1b} show the optical and
infrared color magnitude diagrams of the Blanco 1 cluster from which bona
fide very low mass stars and brown dwarfs are identified.

\section{The mass function of young open clusters}

The present day mass functions (PDMFs) of 3 young clusters were thus derived
from unbiased samples of probable cluster members, taking into account mass
segregation and not attempting to correct for unresolved binaries.

\begin{figure}[ht]
\sidebyside
{\centerline{\psfig{file=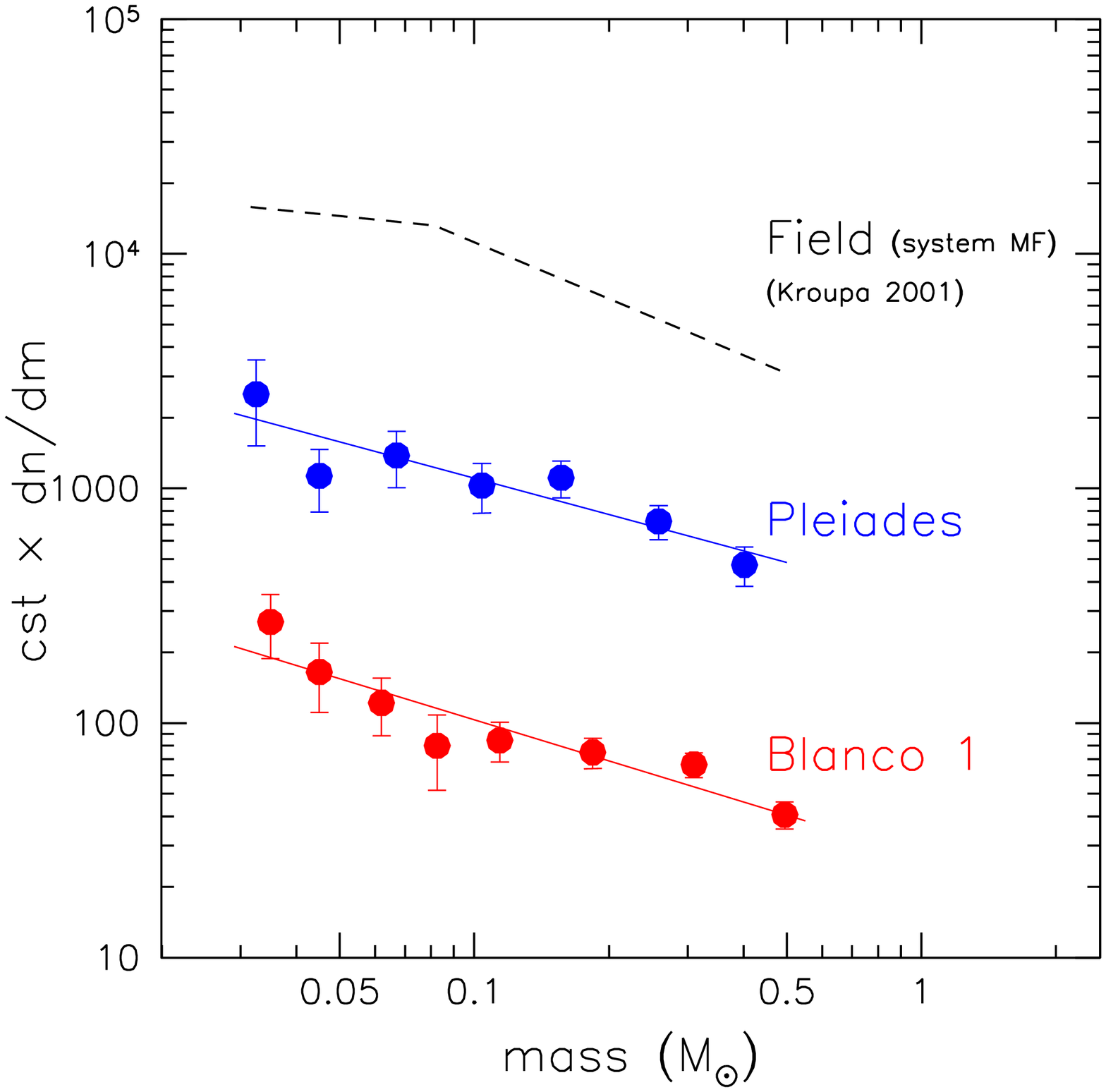,height=2.6in}}
\caption{The mass function of young clusters derived from the CFHT survey in the mass range 0.03-0.50\msunn. The field system MF from Kroupa (2001) is shown for
  comparison ($\alpha_0$=0.2 for $m\leq0.08$\msunn, $\alpha_1$=$0.8$ for
  $0.08<m\leq0.5$\msunn). }}
{\centerline{\psfig{file=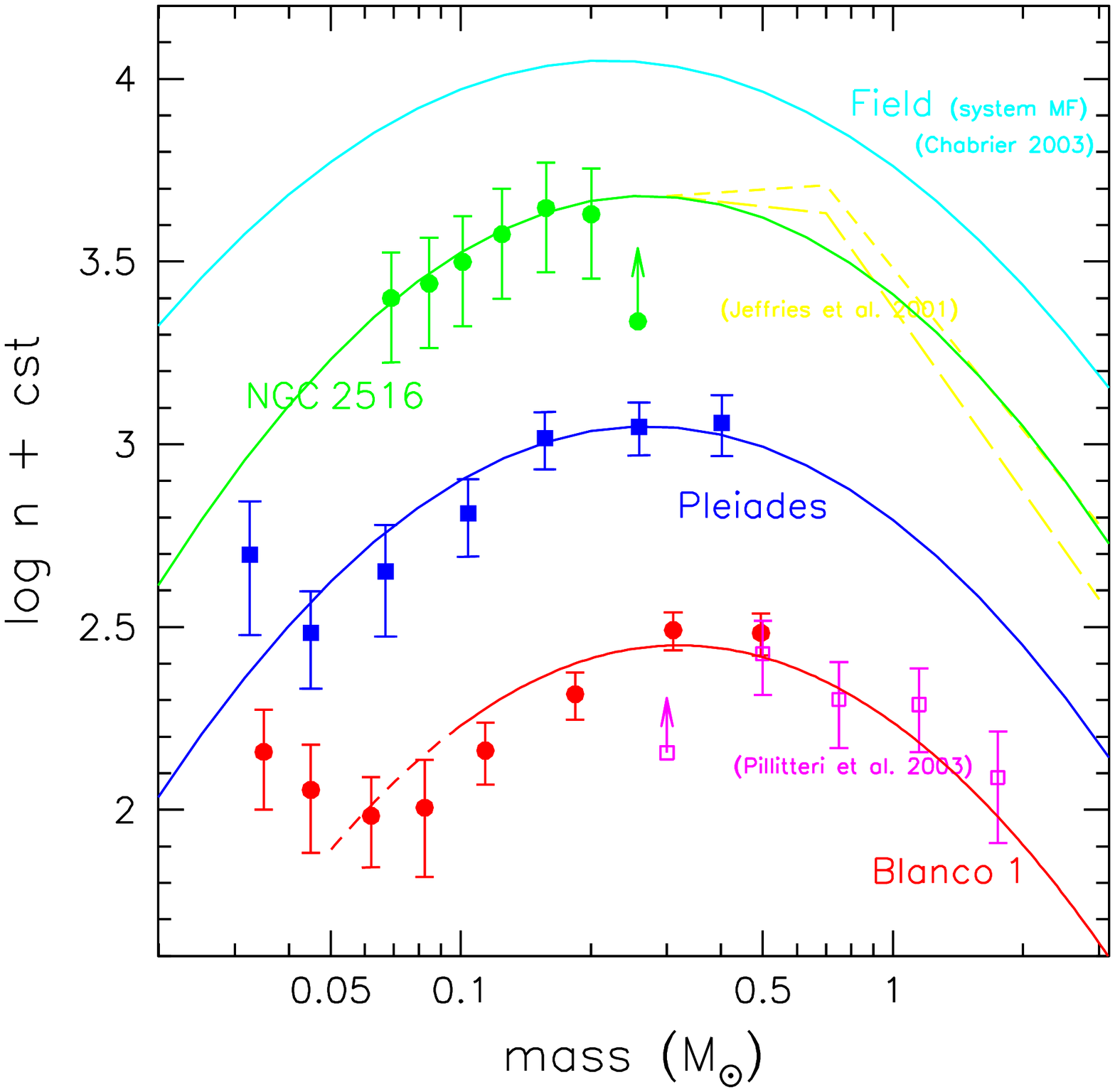,height=2.7in}}
\label{imfpl}
\caption{The mass function of young open clusters, from low mass brown
  dwarfs to the most massive stars. The field system MF from Chabrier
  (2003) is shown for comparison. Parameters of the lognormal functional
  forms are given in Table~2.}}
\label{imflognorm}
\end{figure}

In a Salpeter-like power-law representation, $$\xi(m) = {dn\over dm}\propto m^{-\alpha}$$
we derive $\alpha=0.6\pm0.1$ for the mass distribution of
Pleiades and Blanco 1 members over the 0.03--0.50\msun mass range
(Fig.~\ref{imfpl}). This is broadly consistent with Kroupa's (2001) field
IMF for systems, though we find no evidence for a discontinuity at the
stellar/substellar boundary.

On a broader mass range, from 30 Jupiter masses up to the most massive
stars, the mass function of young clusters is resonably well fitted
(Fig.~\ref{imflognorm}) by a Scalo-like lognormal form :
 \begin{equation}
   \xi_{\rm L}(m) = \frac{dn}{d\log m} \propto \exp \left[ -
     \frac{(\log m - \log m_{0})^2}{2\sigma^{2}} \right]
   \label{eq:plmf}
 \end{equation} 

\noindent with ($m_o, \sigma$) values listed in Table~2. The high mass part of the
MFs in Fig.~\ref{imflognorm} was built from published data (Pillitteri et
al. 2003, Jeffries et al. 2001, Stauffer \& Prosser database, see Moraux et
al. 2003).

\begin{table}[ht]
\caption{Cluster PDMFs lognormal functional fits : ${dn\over d\log m} \propto \exp { {- (\log m - \log m_o)^2} \over {2\sigma^2 } }$}
\begin{tabular*}{\textwidth}{@{\extracolsep{\fill}}lccccc}
\sphline
\it Region &\it $m_o$ &\it $\sigma$ &\it $n_{BD}/n_{tot}$ &\it
$n_\star/n_{tot}$ &\it $\Sigma (m_{BD})/m_{tot}$ \cr
& \it (\msunn) &   & \it & \it  \cr
\sphline
Blanco 1 & 0.32 & 0.50 & 0.11 & 0.89 & 1\% \cr
Pleiades & 0.27 & 0.52 & 0.15 & 0.85 & 1.5\% \cr 
NGC 2516 & 0.27 & 0.51 & 0.15 & 0.85 & 1.5\% \cr
\sphline
Field$^a$  & 0.22 & 0.57 & 0.21 & 0.79 & 2\% \cr
\sphline
\end{tabular*}
\begin{tablenotes}
$^a$ Chabrier's (2003) lognormal galactic {\it system} MF. 
%
\end{tablenotes}
\end{table}

From Fig.~\ref{imflognorm} and Table~2, the PDMF of the 3 young open
clusters appear quite similar, with a characteristic system mass around
0.3\msunn. The fraction of brown dwarfs in each cluster compared to stars
is 10-15\%, which amounts to a substellar mass of $\sim$1\% of the total
cluster mass.  Hence, to first order, the cluster MFs appear invariant
under various probably different initial conditions (stellar density,
metallicity, etc.).

The cluster PDMFs are also similar to the field system MF (Chabrier 2003,
Moraux, Kroupa \& Bouvier 2004).  Small differences may exist, with a
characteristic mass possibly slightly higher in clusters than in the field
and, consequently, a larger BD-to-star fraction in the field. Whether these
differences result from the early dynamical evolution of the cluster, with
a preferential loss of low mass members, or from remaining uncertainties in
the field and clusters MF estimates requires more detailed studies.

\end{document}